\documentclass[preprint]{elsarticle}
\usepackage{textcomp}
\usepackage{amssymb}
\usepackage{amsmath}
\usepackage{graphicx}
\usepackage{hyperref}
\usepackage{pgf}
\usepackage{pdftricks}
\begin{psinputs}

\usepackage{pstricks,pst-fill,pst-grad}

\end{psinputs}
\usepackage{graphicx}
\usepackage{epsfig}
\usepackage{graphics}
\usepackage{pstricks,pst-coil,pst-fill,pst-plot}
\usepackage{epstopdf}
\usepackage{enumerate}

\newtheorem{thm}{Theorem}[section]

\newtheorem{lemma}{Lemma}[section]
\newtheorem{cor}{Corollary}[section]
\newtheorem{Def}{Definition}[section]
\newtheorem{rem}{Remark}[section]

\newcommand{\be}[1]{\begin{equation}\label{#1}}
\newcommand{\ba}[1]{\begin{eqnarray}\label{#1}}
\newcommand{\ee}{\end{equation}}
\newcommand{\ea}{\end{eqnarray}}

\newcommand{\bra}[1]{\langle{#1}|}
\newcommand{\ket}[1]{|{#1}\rangle}
\def\End{\operatorname{End}}

\def\tens{\otimes}

\def\pl{\prod\limits}

\def\build#1_#2^#3{\mathrel{
\mathop{\kern 0pt#1}\limits_{#2}^{#3}}}

\newcommand{\Cset }{{\mathbb C}}

\usepackage{euscript}

\def\la{\lambda}

\def\pl{\prod\limits}
\def\lt({\left(}
\def\rt){\right)}

\def\det{\operatorname{det}}

\def\la{\lambda}

\begin{document}

\begin{frontmatter}
\title
{Elliptic dynamical reflection algebra and partition function of SOS model with reflecting end  }
\author{G.Filali}
\address{Universit\'e
de Cergy-Pontoise, LPTM UMR 8089 du CNRS,  2 av. Adolphe Chauvin, 95302 Cergy-Pontoise, France, e-mail: ghali.filali@u-cergy.fr}

\begin{abstract}
  We introduce in this paper an elliptic dynamical reflection algebra describing an SOS model with reflecting end. Using factorizing Drinfel'd twist, we compute the partition function of this model with domain wall boundary conditions. We show that it can be represented in the form of a single Izergin determinant.

\end{abstract}

\end{frontmatter}

\section{Introduction}
SOS type models in statistical mechanics play an important role for mathematical physics. Since the pioneer works of Baxter on the eight vertex model \cite{Bax73}, they arise as a necessary step towards resolution of vertex model without charge conservation \cite{FadST79,Fel06b}. As they are described by dynamical Yang-Baxter algebra, they also arise as underlining algebraic structure of conformal field theories \cite{Fel06a}. From the algebraic point of view, they are intensively studied due to their relations to quasi-hopf structures \cite{Enr98} and current algebras \cite{JKMO99,PakRS08}. Furthermore, they are related to combinatorics and dynamical enumeration of alternating sign matrix \cite{Ros09}.

Within the Quantum Inverse Scattering frameworks, partition functions are essential for the study of correlations functions \cite{KitMT99,KitMT00,KitKMNST08,KitKMNST07}. The partition function for such elliptic SOS model with domain wall boundary conditions has already been computed by different methods in \cite{PakRS08,Ros09}, but due to the dynamical nature of the underlining algebras, it is presented as a sum of matrix determinants. Recently, partition function of an  alternative trigonometric SOS model with reflecting end has been computed as a single determinant \cite{FilK10}. It is shown there that such boundaries permit to avoid the difficulty of the inherent dynamical algebra, by means of a dynamical reflection algebra. We generalize in this letter the result to the elliptic case, which is the most general SOS models, by use of the concept of factorizing Drinfel'd twist.
\section{Preliminaries: elliptic theta functions}

Let $\tau$ be a fixed complex parameter such that: $Im(\tau)>0$ and denote $p=e^{2i\pi\tau}$. Throughout this paper, we will use the notation
\begin{equation}
h(\lambda)=e^{\lambda}\prod_{i=0}^{\infty}(1-p^{i}e^{-2\lambda})(1-p^{i+1}e^{2\lambda}).
\end{equation}
Up to a multiplicative factor, $h(\lambda)$ equals the Jacobi theta function $\theta_{1}(i \lambda)$ \cite{Whitaker}.
This function is odd and satisfy the addition rule
\begin{align}
h(x+u)h(x-u)&h(y+v)h(y-v)-h(x+v)h(x-v)h(y+u)h(y-u)\\ \nonumber
&=h(x+y)h(x-y)h(u+v)h(u-v).
\end{align}
In the degenerate case, we have: $\lim_{p\rightarrow0}h(\lambda)= 2\sinh(\lambda)$.

\begin{Def}
$f$ is a theta function of norm $t$ and order $N$ if there exist $N$ constants $\{\xi_{i}\}_{i=1,...,N}$ and $\sum_{i=1}^{N}\xi_{i}=t$ such that
\begin{equation}
f(\lambda)=\prod_{i=1}^{N}h(\lambda+\xi_{i}).
\end{equation}
\end{Def}
Using this definition, we have the classical theorem \cite{Weber}
\begin{thm}
Let $f$ be
\begin{equation}
f(\lambda)=\sum_{j}\prod_{i=1}^{N}h(\lambda+\xi^{j}_{i}),
\end{equation}
where for any $j$, $\sum_{i=1}^{N}\xi^{j}_{i}=t$. Then $f$ is a theta function of order $N$ and norm $t$.
\end{thm}
This theorem gives a powerful tool to prove theta functions identities.
\section{Algebraic framework}
We start by introducing the dynamical reflection algebra which is underlining integrability of the elliptic SOS model with reflecting end as we will show in this paper. This algebra is actually built as a comodule over the elliptic quantum group $E_{\tau,\eta}(sl_{2})$. The main object for defining the elliptic quantum group $E_{\tau,\eta}(sl_{2})$ \cite{Fel06a} is the dynamical $R$ matrix, $R:\Cset \times \Cset \longrightarrow \End(V\tens V)$, $V\sim\Cset^2$
\begin{equation}
R(\lambda;\theta)=
\begin{pmatrix}
R^{+ +}_{+ +}(\lambda;\theta)&0&0&0\\
0 & R^{+ -}_{+ -}(\lambda;\theta)& R^{+ -}_{ - +}(\lambda;\theta) & 0\\
0 &R^{ - +}_{+ -}(\lambda;\theta) & R^{- +}_{- +}(\lambda;\theta) & 0\\
0 & 0 & 0 & R^{- -}_{- -}(\lambda;\theta)\\
\end{pmatrix},
\end{equation}

which satisfies the dynamical Yang-Baxter equation

\begin{align}\label{DYBE}
R_{12}(\lambda_{1}-\lambda_{2};\theta-\eta\sigma^{z}_{3})&R_{13}(\lambda_{1}-\lambda_{3};\theta)R_{23}(\lambda_{2}-\lambda_{3};\theta-\eta\sigma^{z}_{1})\nonumber \\=&R_{23}(\lambda_{2}-\lambda_{3};\theta)R_{13}(\lambda_{1}-\lambda_{3};\theta-\eta\sigma^{z}_{2})R_{12}(\lambda_{1}-\lambda_{2};\theta),
\end{align}
where we denote by $\sigma_{a}^{x,y,z}$ the usual Pauli matrices in the two dimensional space $V_{a}\sim\Cset^{2}$. We are interested here in the elliptic solution of this equation \cite{JKMO88}, which is the most general case

\begin{align}\label{dynamicalRmatrix}
R^{+ +}_{+ +}(\lambda;\theta)&=R^{- -}_{- -}(\lambda;\theta)=h(\lambda+\eta) \nonumber\\
R^{+ -}_{+ -}(\lambda;\theta)&=R^{- +}_{ - +}(\lambda;-\theta)=\frac{h(\lambda) h(\theta-\eta)}{h(\theta)} \\
R^{+ -}_{ - +}(\lambda;\theta)&=R^{- +}_{ + - }(\lambda;-\theta)= \frac{h(\eta) h(\theta-\lambda)}{h(\theta)},\nonumber
\end{align}
The elliptic $E_{\tau,\eta}(sl_{2})$ quantum group is the algebra generated by meromorphic funtions of the generator of $\mathfrak{h}$, the Cartan subalgebra of $sl_{2}$, that we denote by $\sigma^{z}$ , and the matrix elements of $T(\lambda;\theta)=
\begin{pmatrix}
A(\lambda;\theta)&B(\lambda;\theta)\\
C(\lambda;\theta)&D(\lambda;\theta)\\
\end{pmatrix}
 \in \End(\Cset^{2})$ with non-commutative entries, satisfying the dynamical Yang-Baxter algebra relations

 \begin{align}\label{DYBA}
R_{12}(\lambda_{1}-\lambda_{2};\theta-\eta\sigma^{z})&T_{1}(\lambda_{1};\theta)T_{2}(\lambda_{2};\theta-\eta\sigma^{z}_{1})\nonumber\\=&T_{2}(\lambda_{2};\theta)T_{1}(\lambda_{1};\theta-\eta\sigma^{z}_{2})R_{12}(\lambda_{1}-\lambda_{2};\theta).
\end{align}

We are interested here only in diagonalizable $\mathfrak{h}$-module $V$ where the weight zero property holds

\begin{equation}
[T_{0}(\lambda;\theta),\sigma^{z}_{0}+\sigma^{z}_{V}]=0.
\end{equation}

We define the following Dynamical Reflection Algebra generated by meromorphic functions of $\sigma^{z} \in \mathfrak{h}$ and the matrix element of $\mathcal{T}(\lambda;\theta) \in End(\mathbb{C}^{2})$ with non commutative entries subject to the relations

\begin{align}\label{DRA}
R_{12}(\lambda_{1}-\lambda_{2};\theta-\eta\sigma^{z})&\mathcal{T}_{1}(\lambda_{1};\theta)R_{21}(\lambda_{1}+\lambda_{2};\theta-\eta\sigma^{z})\mathcal{T}_{2}(\lambda_{2};\theta)\\\nonumber
=\mathcal{T}_{2}(\lambda_{2};\theta)&R_{12}(\lambda_{1}+\lambda_{2};\theta-\eta\sigma^{z})\mathcal{T}_{1}(\lambda_{1};\theta)R_{21}(\lambda_{1}-\lambda_{2};\theta-\eta\sigma^{z}).
\end{align}

Let $\mathcal{K}:\mathbb{C}\times\mathbb{C}\longrightarrow End(\mathbb{C}^{2})$, be a (scalar) representation of this algebra in $\mathbb{C}$ (i.e $\mathbb{C}$-number matrix) viewed as one dimensional $\mathfrak{h}$-module of $sl_{2}$ with the standard action on $v\in\mathbb{C},\ \  \sigma^{z}.v=0$

\begin{align}\label{DYRE}
R_{12}(\lambda_{1}-\lambda_{2};\theta)&\mathcal{K}_{1}(\lambda_{1};\theta)R_{21}(\lambda_{1}+\lambda_{2};\theta)\mathcal{K}_{2}(\lambda_{2};\theta)\nonumber\\=&\mathcal{K}_{2}(\lambda_{2};\theta)R_{12}(\lambda_{1}+\lambda_{2};\theta)\mathcal{K}_{1}(\lambda_{1};\theta)R_{21}(\lambda_{1}-\lambda_{2};\theta).
\end{align}

This is essentially the reflection equation introduced in \cite{Skl88}, with the dynamical $R$-matrix instead of the usual one. A representation as above is said of weight zero if
\begin{equation}
[\mathcal{K}_{0}(\lambda;\theta),\sigma^{z}_{0}].
\end{equation}

This means that $\mathcal{K}$ is diagonal solution of the above equation. Let $T(\lambda;\theta)$ a weight zero representation of the dynamical Yang-Baxter algebra in $V$. Then
\begin{equation}
\mathcal{T}(\lambda;\theta)=T(\lambda,\theta)\mathcal{K}(\lambda;\theta)T^{-1}(-\lambda;\theta),
\end{equation}
is a weight zero representation of the dynamical reflection algebra in $\mathbb{C}\otimes V$.

In this letter, we are mostly interested in a particular representation of this dynamical reflection algebra, which is built on the well-known evaluation representation of $E_{\tau,\eta}(sl_{2})$ in the space $V=\otimes_{i=1}^{N}\mathbb{C}_{i}^{2}$ which is constructed from the $R$-matrix \eqref{dynamicalRmatrix}

\begin{align}\label{bulkopera}
T_{0}(\lambda;\theta)&=R_{01}(\lambda-\xi_{1};\theta-\eta\sum_{i=2}^{N}\sigma_{i}^{z})...R_{0N}(\lambda-\xi_{N};\theta)\\\nonumber
&=\begin{pmatrix}
A(\lambda;\theta)&B(\lambda;\theta)\\
C(\lambda;\theta)&D(\lambda;\theta)\\
\end{pmatrix},
\end{align}

and the specific diagonal solution of \eqref{DYRE}

\begin{equation}\label{boundaryK}
\mathcal{K}(\lambda;\theta)=
\begin{pmatrix}
\frac{h(\theta+\zeta-\lambda)}{h(\theta+\zeta+\lambda)}&0\\
0&\frac{h(\zeta-\lambda)}{h(\zeta+\lambda)}
\end{pmatrix},
\end{equation}which depends on an arbitrary parameter $\zeta$. So our main object of study is the boundary monodromy matrix, a representation of \eqref{DRA} in $\Cset \times V$, $V=\otimes_{i=1}^{N}\mathbb{C}_{i}^{2}$

\begin{align}\label{DDROW}
\mathcal{T}(\lambda;\theta)&=\widehat{\gamma}(\lambda)T(\lambda;\theta)\mathcal{K}(\lambda;\theta)T^{-1}(-\lambda;\theta)\\ \nonumber
&=\begin{pmatrix}
\mathcal{A}(\lambda;\theta)&\mathcal{B}(\lambda;\theta)\\
\mathcal{C}(\lambda;\theta)&\mathcal{D}(\lambda;\theta)\\
\end{pmatrix},
\end{align} with  normalization coefficients
\begin{equation}
\widehat{\gamma}(\lambda)=(-1)^{N}\prod_{i=1}^{N}h(\lambda+\xi_{i}-\eta)h(\lambda+\xi_{i}+\eta).
\end{equation}

This representation has a clear statistical mechanics interpretation, it describes an elliptic SOS model with reflecting end. Before introducing this statistical mechanics model, we give in the next section convenient expression for the boundary monodromy operators \eqref{DDROW}.

%\widehat{T}(\lambda;\theta)\nonumber\\
%=&R_{01}(\lambda-\xi_{1};\theta-\eta\sum_{i=2}^{N}\sigma_{i}^{z})...R_{0N}(\lambda-\xi_{N};\theta)\nonumber\\ %\times&\mathcal{K}(\lambda;\theta)R_{N0}(\lambda+\xi_{N};\theta)...R_{10}(\lambda+\xi_{1};\theta-\eta\sum_{i=2}^{N}\sigma_{i}^{z})
%\end{align}

\section{Symmetry and Drinfel'd twist}
In this section, we collect some useful relations of the $R$-matrix and corresponding monodromy operators. This relations will help us to obtain manageable expressions of the boundary operators \eqref{DDROW} with the help of the factorizing Drinfel'd twist.

\subsection{$R$-matrix and boundary operators}
The $R$-matrix satisfies three important properties

{ 1.  Zero weight\footnote{This property is sometimes referred in statistical mechanics as the Ice Rule}}:
\medskip
\begin{equation}\label{weightzero}
[\sigma_{1}^{z}+\sigma_{2}^{z},R_{12}(\lambda;\theta)]=0
\end{equation}

This symmetry reflects the six vertex texture of the statistical weights: $R_{\alpha \beta}^{\mu \nu}=0$ unless $\alpha+\beta=\mu+\nu$. It is easy to see that this relation induces a similar relations for the transposed $R$ matrix
\begin{equation}\label{transposedweightzero}
[\sigma_{1}^{z}-\sigma_{2}^{z},R^{t_{1}}_{12}(\lambda;\theta)]=0
\end{equation}

{ 2. Unitarity}:

\medskip

\begin{equation}
R_{12}(\lambda;\theta)R_{21}(-\lambda;\theta)=-h(\lambda-\eta)h(\lambda+\eta)Id
\end{equation}

\medskip
{ 3. Crossing Symmetry}:

\medskip

We write the crossing relation for the dynamical $R$-matrix in the following compact form:
\begin{equation}\label{crossR}
-\sigma_{1}^{y}:R_{12}^{t_{1}}(-\lambda-\eta;\theta+\eta\sigma_{1}^{z}):\sigma_{1}^{y}
\frac{h(\theta-\eta\sigma_{2}^{z})}{h(\theta)}
=R_{21}(\lambda;\theta)
\end{equation} where we assume the following normal ordering : the $\sigma_{1}^{z}$ in the argument of the $R$-matrix (which does not commute with it) is always on the right of all other operators involved in the definition of $R$.

\subsection{Boundary monodromy matrix}

Using the crossing relation \eqref{crossR} and the zero weight for the transposed $R$-matrix \eqref{transposedweightzero} we  get
\begin{align}\label{Tcross}
\widehat{\gamma}(\lambda)T^{-1}(-\lambda;\theta)\equiv & R_{N0}(\lambda+\xi_{N};\theta)...R_{10}(\lambda+\xi_{1};\theta-\eta\sum_{i=2}^{N}\sigma_{i}^{z})\nonumber\\
=&\gamma(\lambda)\sigma_{0}^{y}T^{t_{0}}(-\lambda-\eta;\theta+\eta\sigma_{0}^{z})\sigma_{0}^{y}\nonumber\\
&\qquad\qquad\quad
\frac{h(\theta-\eta\mathbf{S}^z)}{h(\theta)},
\end{align}
with  normalization coefficients
\begin{equation}
\gamma(\lambda)=(-1)^{N},
\end{equation}
and the short notation: $\sigma^{z}_{V}=\mathbf{S}^z=\sum_{i=1}^{N}\sigma_{i}^{z}$. A very important decomposition of the $\mathcal{B}$ operators is given by means of the generalized crossing relation for the bulk monodromy matrix \eqref{Tcross}.
It implies for the $\mathcal{B}$ operators
\begin{align}\label{BoundaryBulk}
\mathcal{B}(\lambda;\theta)=&\gamma(\lambda)\Big(\!\mathcal{K}^{-}_{-}B(\lambda;\theta)A(-\lambda-\eta;\theta+\eta)-\mathcal{K}^{+}_{+}A(\lambda;\theta)B(-\lambda-\eta;\theta-\eta)\!\Big)\nonumber\\
&\times \frac{h(\theta-\eta\mathbf{S}^z)}{h(\theta)}.
\end{align}

Furthermore, the equation \eqref{DRA} contains the commutation  relations for the generators: $\mathcal{A}(\lambda;\theta)$, $\mathcal{B}(\lambda;\theta)$, $\mathcal{C}(\lambda;\theta)$ and $\mathcal{D}(\lambda;\theta)$. The only  one which is important for  the computation of the partition function is the relation for the $\mathcal{B}$ operators
\begin{equation}
\label{commutB}
\mathcal{B}(\lambda_{1};\theta)\mathcal{B}(\lambda_{2};\theta)=\mathcal{B}(\lambda_{2};\theta)\mathcal{B}(\lambda_{1};\theta)
\end{equation}

\subsection{Symmetric representation and the F-basis}
Representation of Drinfel'd twist \cite{Drin90} was first applied by Maillet and Sanchez de Santos \cite{Mai96} in order to obtain completely symmetric representation of the bulk monodromy operators for Yang-Baxter type algebra, which are highly non local in terms of the quantum local operators. The idea is to perform a change of basis in the space of states where the bulk monodromy operators remains completely symmetric.

This representation is very useful for the computation of partition functions, scalar products and correlation functions of integrable spin chains with periodic boundary conditions \cite{KitMT99,KitMT00} but also for diagonal boundary conditions \cite{KitKMNST07,KitKMNST08}, as it reduces drastically the combinatorial difficulty of handling highly non-local representation.

In this representation, not only the bulk monodromy operators remain symmetric, we also have a direct insight into the analytical property of this operators.

For the evaluation representation of dynamical Yang-Baxter algebras, such representation of the bulk monodromy operators was proposed in \cite{Al00}. This construction is based on a dynamical $F$-matrix which factorizes the dynamical $R$-matrix in the following way
\begin{equation}
F_{21}(-\lambda;\theta)R_{12}(\lambda;\theta)=F_{12}(\lambda;\theta).
\end{equation}
After a suitable co-product over all quantum spaces it leads to a change of basis $F_{\{\xi\}}$ where the bulk operators $A(\lambda;\theta),B(\lambda;\theta)$ \eqref{bulkopera} have symmetric expressions

\begin{align}
\overline{A}(\lambda;\theta)&=F_{\{\xi\}}(\theta)A(\lambda;\theta)F^{-1}_{\{\xi\}}(\theta-\eta)\\ \nonumber
&=\frac{h(\theta-\eta)}{h\left(\theta+\eta(\frac{N-\mathbf{S}^{z}}{2}-1)\right)}\otimes_{i=1}^{N}
\begin{pmatrix}
h(\lambda-\xi_{i}+\eta)&0\\
0&h(\lambda-\xi_{i})\\
\end{pmatrix}_{i},
\end{align}

and

\begin{align}
\overline{B}(\lambda;\theta)&=F_{\{\xi\}}(\theta)B(\lambda;\theta)F^{-1}_{\{\xi\}}(\theta+\eta)\\ \nonumber
&=\frac{h(\eta)}{h(\theta)}\sum_{i=1}^{N}h(\theta-\lambda+\xi_{i})\sigma_{i}^{-}\otimes_{j\neq i}^{N}
\begin{pmatrix}
h(\lambda-\xi_{j}+\eta)&0\\
0&\frac{h(\lambda-\xi_{i})h(\xi_{i}-\xi_{j}+\eta)}{h(\xi_{i}-\xi_{j})}\\
\end{pmatrix}_{j}.
\end{align}

A very important property is that the reference states $\ket{0}=\prod_{i=1}^{N} \uparrow_{\xi_{i}}$ and $\ket{\bar{0}}=\prod_{i=1}^{N} \downarrow_{\xi_{i}}$ are left and right invariant under the action of $F_{\{\xi\}}$
\begin{equation}
F_{\{\xi\}}(\theta)\ket{0}=F^{-1}_{\{\xi\}}(\theta)\ket{0}=\ket{0},\bra{0}F_{\{\xi\}}(\theta)=\bra{0}F^{-1}_{\{\xi\}}(\theta)=\bra{0},
\end{equation}
and
\begin{equation}
F_{\{\xi\}}(\theta)\ket{\bar{0}}=F^{-1}_{\{\xi\}}(\theta)\ket{\bar{0}}=\bra{\bar{0}},\bra{\bar{0}}F_{\{\xi\}}(\theta)=\bra{\bar{0}}F^{-1}_{\{\xi\}}(\theta)=\bra{\bar{0}}.
\end{equation}

Using the decomposition \eqref{BoundaryBulk} of the boundary operator $\mathcal{B}(\lambda;\theta)$, it is easy to compute its expression in this new basis

\begin{align}\label{Bsymmetric}
\overline{\mathcal{B}}(\lambda;\theta)&=F_{\{\xi\}}(\theta)\mathcal{B}(\lambda;\theta)F^{-1}_{\{\xi\}}(\theta)\\ \nonumber
&=\gamma(\lambda)\sum_{i=1}^{N}\Big\{\frac{h(\theta+\zeta+\xi_i)}{h(\theta+\zeta+\lambda)}\frac{h(\zeta-\xi_i)}
{h(\zeta+\lambda)}h(2\lambda)h(\eta)\\\nonumber
&\sigma_{i}^{-}\otimes_{j\neq i}^{N}
\begin{pmatrix}
h(\lambda+\xi_{j})h(\lambda-\xi_{j}+\eta)&0\\\nonumber
0&\frac{h(\lambda-\xi_{j})h(\lambda+\xi_{j}+\eta)h(\xi_{i}-\xi_{j}+\eta)}{h(\xi_{i}-\xi_{j})}\\\nonumber
\end{pmatrix}_{j}\Big\}\\\nonumber
&\times\frac{h(\theta-\eta\mathbf{S}^z)}{h(\theta+\eta\frac{N-\mathbf{S}^z}{2})}.
\end{align}

\section{Elliptic SOS model with reflecting end}
Let us now introduce the SOS model, which is a two-dimensional statistical mechanics lattice model where Boltzmann weights are attached to each \textit{face}. There are 6 possible face configurations where each \textit{height} $\theta$ can differ only by $\pm \eta$ for adjacent sides

\[
\begin{array}{ccc}\begin{array}{ccc}
\vphantom{\sum\limits_1^2}\theta -\eta &\vline&\theta -2\eta\\
\hline
\vphantom{\sum\limits_1^2}\theta &\vline&\theta-\eta
\end{array}\qquad&\qquad
\begin{array}{ccc}
\vphantom{\sum\limits_1^2}\theta +\eta&\vline&\theta +2\eta\\
\hline
\vphantom{\sum\limits_1^2}\theta &\vline&\theta +\eta
\end{array}
\qquad&\qquad
\begin{array}{ccc}
\vphantom{\sum\limits_1^2}\theta -\eta&\vline&\theta \\
\hline
\vphantom{\sum\limits_1^2}\theta &\vline&\theta +\eta
\end{array}
\\
\bigskip
\\
\begin{array}{ccc}
\vphantom{\sum\limits_1^2}\theta +\eta&\vline&\theta \\
\hline
\vphantom{\sum\limits_1^2}\theta &\vline&\theta-\eta
\end{array}\qquad&\qquad
\begin{array}{ccc}
\vphantom{\sum\limits_1^2}\theta +\eta&\vline&\theta \\
\hline
\vphantom{\sum\limits_1^2}\theta &\vline&\theta +\eta
\end{array}
\qquad&\qquad
\begin{array}{ccc}
\vphantom{\sum\limits_1^2}\theta -\eta&\vline&\theta \\
\hline
\vphantom{\sum\limits_1^2}\theta &\vline&\theta -\eta
\end{array}
\\

\end{array}
\]

The corresponding statistical weights $R^{a b}_{c d}$ are collected into the dynamical $R$ matrix \eqref{dynamicalRmatrix}. In the limit $p \rightarrow 0$, the model becomes equivalent to the trigonometric SOS model. Furthermore, the limit $\theta \rightarrow \infty$ gives rise to the well-known six vertex model \cite{Kor82}. Dealing with boundary  weights requires to introduce the boundary matrix $\mathcal{K}(\lambda;\theta)$ which we choose to be \eqref{boundaryK}. Notice that in the trigonometric case, $\mathcal{K}(\lambda;\theta)$ reduces to the trigonometric solution of \cite{FilK10}. We consider this model with a reflecting end, which means that each horizontal line makes a U-turn on the left side of the lattice. As we choose a diagonal solution, it produces two configurations characterized by the weights  $\mathcal{K}_\pm^\pm (\lambda;\theta)$.
\vskip 0.5cm
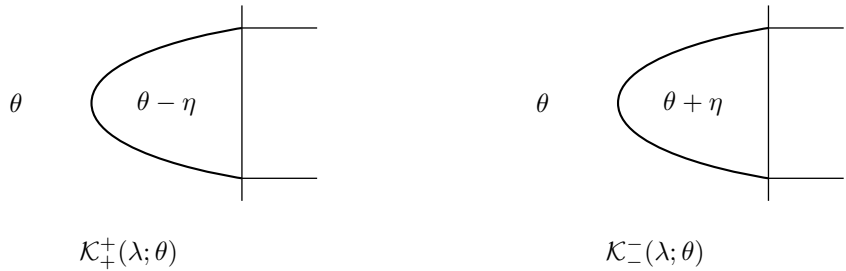
\begin{figure}
\begin{center}
\setlength{\unitlength}{2cm}
\begin{pspicture}(12.5,3)
\psline[linewidth=0.5pt](4,1.5)(5,1.5)
\psline[linewidth=0.5pt](4,3.5)(5,3.5)
\psline[linewidth=0.5pt](4,1.2)(4,3.8)
\pscurve(4,1.5)(2,2.5)(4,3.5)
\rput(3,2.5){$\theta-\eta$}
\rput(1,2.5){$\theta$}
\rput(2.5,0.5){$\mathcal{K}_{+}^+(\lambda; \theta)$}

\psline[linewidth=0.5pt](11,1.5)(12,1.5)
\psline[linewidth=0.5pt](11,3.5)(12,3.5)
\psline[linewidth=0.5pt](11,1.2)(11,3.8)
\pscurve(11,1.5)(9,2.5)(11,3.5)
\rput(10,2.5){$\theta+\eta$}
\rput(8,2.5){$\theta$}
\rput(9.5,0.5){$\mathcal{K}_{- }^-(\lambda; \theta)$}
\end{pspicture}

\caption{Boundary configuration with external height $\theta$.}
\end{center}
\end{figure}
It is important to note that such reflecting end imposes a constant external height $\theta$ for the left side of the lattice. This reflecting end leads to different parametrizations of the weight if they are in the two different half the rows. Indeed, parametrization should respect row and line multiplication, and also some fundamental symmetry of the $R$ matrix (as the zero weight \eqref{weightzero}). We can easily check that this convention leads to a well defined inhomogeneous model:
\vskip 0.5cm

\begin{figure}[h]
\begin{center}
$ R^{c-d, a-c}_{a-b, b-d}(\lambda_{j}+\xi_{i};a)$
\end{center}
\begin{center}
\includegraphics[width=2.5cm]{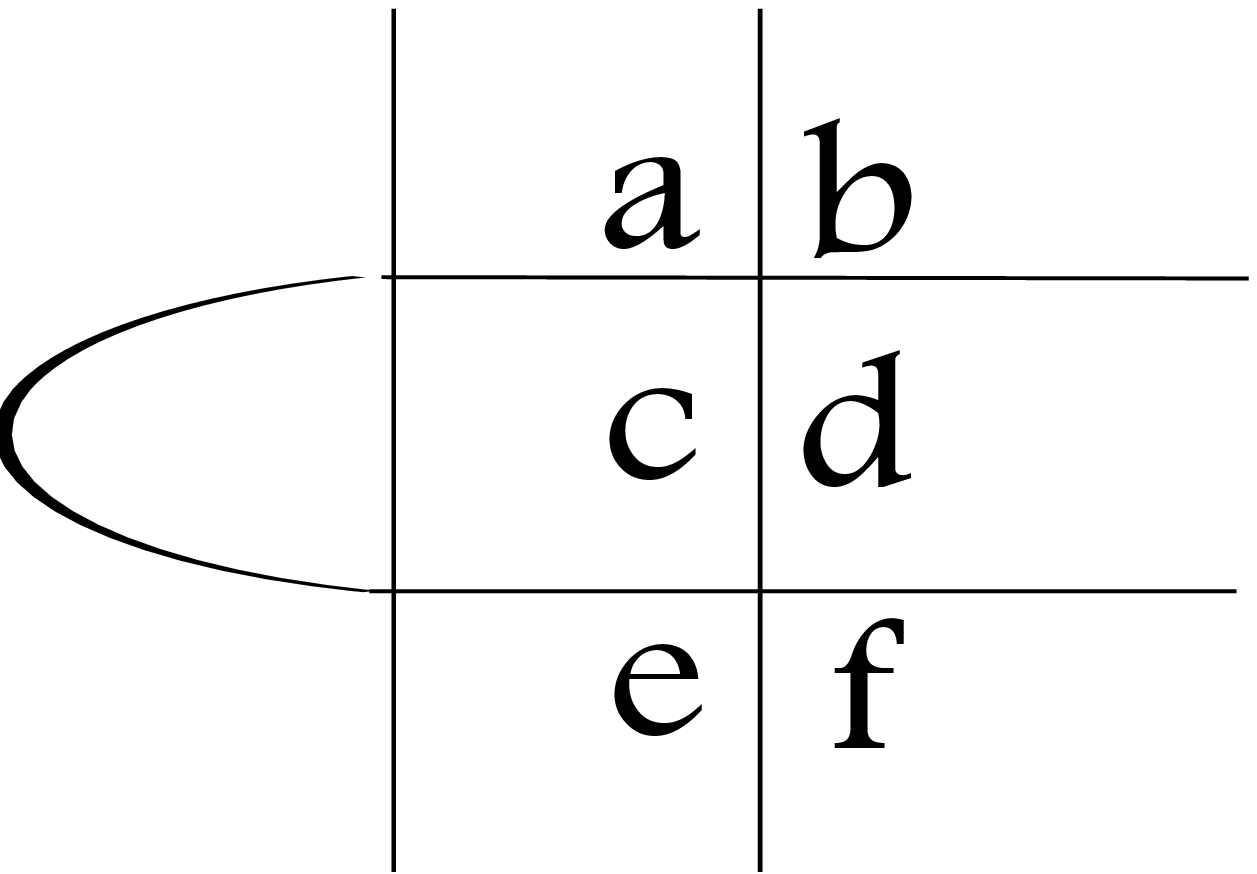}
\end{center}
\begin{center}
$R^{f-d, e-f}_{e-c, c-d}(\lambda_{j}-\xi_{i};e)$
\end{center}
\end{figure}

\vskip 0.5cm

Here we consider the model with \textit{domain wall boundary conditions}, the heights decrease from left to right on the upper boundary, the heights grow from left to right on the lower boundary. As left external height is fixed these two conditions determine completely the configuration on the right boundary (heights decreasing in the upward direction).

\begin{figure}[ht]
\begin{center}
\includegraphics[width=4.5cm]{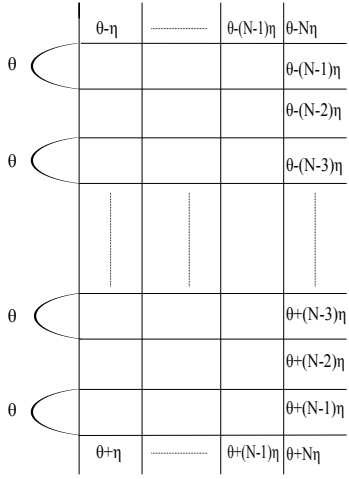}
\end{center}
\begin{center}
\caption{Domain Wall Boundary Conditions}
\end{center}
\end{figure}

\section{Partition Function}
The partition function of the SOS model introduced in the previous section
can be written in terms of the boundary monodromy matrix
\begin{align}\label{PartFun}
Z_{N,2N}(\{\lambda\},\{\xi\},\theta)=&\prod_{i=1}^{N} \uparrow_{\lambda_{i}}\prod_{j=1}^{N} \downarrow_{\xi_{j}}\prod_{i=1}^{N}   \mathcal{T}(\lambda_{i};\theta)
\prod_{i=1}^{N} \uparrow_{\xi_{i}}\prod_{j=1}^{N} \downarrow_{\lambda_{j}}\\ \nonumber
=&
\bra{\bar{0}}\prod_{i=1}^{N}
\mathcal{B}(\lambda_{i};\theta)  \ket{0}\\ \nonumber
=&\bra{\bar{0}}\prod_{i=1}^{N}
\mathcal{\overline{B}}(\lambda_{i};\theta)  \ket{0}
\end{align}
 We follow the standard way to compute the partition function \cite{Kor82,Ize87}, first we establish a set of properties defining it in an unique way  and then we propose a determinant formula which satisfies all these conditions.

The partition function \eqref{PartFun} satisfies the following  properties:

\begin{enumerate}[i)]
\item $Z_{N,2N}(\{\lambda\},\{\xi\},\theta)$ is symmetric in $\lambda_{i}$.
This property follows from the commutation relation (\ref{commutB}) for the operators $\mathcal{B}$.
\item $Z_{N,2N}(\{\lambda\},\{\xi\},\theta)$ is symmetric in $\xi_{i}$.
This is a direct consequence of the Dynamical Yang-Baxter Equation \eqref{DYBE}.\\ It is sufficient to insert $R_{i+1,i}(\xi_{i+1}-\xi_{i};\theta-\eta\sum_{j=i+2}^{N})$ in  (\ref{PartFun}) to get the symmetry for any elementary permutation ${\xi_{i}\leftrightarrow\xi_{i+1}}$.
\item For each  parameter $\lambda_{i}$ the normalized partition function
\begin{equation}\label{normPartFun}
\tilde{Z}_{N,2N}(\{\lambda\},\{\xi\},\theta)=\prod_{i=1}^{N}\frac{h(\theta+\zeta+\lambda_i)h(\zeta+\lambda_i)}{h(2\lambda_i)}Z_{N,2N}(\{\lambda\},\{\xi\},\theta),
\end{equation}
is a theta function of order $2N-2$ and norm $(N-1)\eta$ with respect to the variable $\lambda_{i}$. To prove it, we use the representation of the partition function in the F-basis and  consider the action of the most right $\overline{\mathcal{B}}(\lambda_{N};\theta)$ operator. Due to the symmetric representation of $\overline{\mathcal{B}}(\lambda_{N};\theta)$ in the $F$-basis, the operator $$\frac{h(\theta+\zeta+\lambda_N)h(\zeta+\lambda_N)}{h(2\lambda_N)}\overline{\mathcal{B}}(\lambda_{N};\theta)$$ acts due to \eqref{Bsymmetric} as
$$\sum_{i=1}^{N} \prod_{k=1,k \neq i}^{N} h(\lambda_{N}+\xi_{k})h(\lambda_{N}-\xi_{k}+\eta)$$
 which is a theta function of the desired form thanks to the standard theorem of Section 1. This property remains true for any $\lambda_{i}$ with the help of i).
\item For $N=1$ the partition function is just a sum of two terms
\begin{multline}
Z_{1,2}(\lambda,\xi,\theta)=\frac{h(\eta) h(\theta-\eta)}{h^2(\theta)}\\
\times\left(\frac{h(\theta+\zeta-\lambda)}{h(\theta+\zeta+\lambda)}h(\lambda-\xi)h(\theta+\lambda+\xi)\right.\\
\left.+\frac{h(\zeta-\lambda)}{h(\zeta+\lambda)}h(\lambda+\xi)h(\theta-\lambda+\xi)\right)
\end{multline}
as there are only two configurations possible.
\item Recursive relations.\\
  There are two points where we can easily establish recursive relations, fixing the configuration in the lower right or the upper right corner by setting  $\lambda_{1}=\xi_{1}$ or  $\lambda_{N}=-\xi_{1}$. It is easy to see that it leads to the following recursive relations
\begin{align}
Z_{N,2N}&\left. (\{\lambda\},\{\xi\},\{\theta\})\vphantom{\prod_{i=1}}\right|_{\la_1=\xi_1}=
\frac{h(\eta)h(\zeta-\lambda_1)}{h(\zeta+\lambda_1)}
\nonumber\\
&\times\prod_{i=1}^{N}h(\lambda_{i}+\xi_{1})\frac{h(\theta+(N-2i)\eta)}{h(\theta+(N-2i+1)\eta)}\nonumber\\
&\times\prod_{i=2}^{N}h(\lambda_{1}-\xi_{i}+\eta)h(\lambda_{1}+\xi_{i}+\eta)h(\lambda_{i}-\xi_{1}+\eta)
\nonumber\\ &\times Z_{(N-1),2(N-1)}(\{\lambda\}_{2\dots N},\{\xi\}_{2\dots N},\{\theta\})
\end{align}
\begin{align}
Z_{N,2N}&\left. (\{\lambda\},\{\xi\},\{\theta\})\vphantom{\prod_{i=1}}\right|_{\la_N=-\xi_1}=
\frac{h(\eta)h(\theta+\zeta-\lambda_N)}{h(\theta+\zeta+\lambda_N)}
\nonumber\\
&\times\prod_{i=1}^{N}h(\lambda_{i}-\xi_{1})\frac{h(\theta+(N-2i)\eta)}{h(\theta+(N-2i+1)\eta)}\nonumber\\
&\times\prod_{i=2}^{N}h(\lambda_{N}+\xi_{i}+\eta)h(\lambda_{N}-\xi_{i}+\eta)h(\lambda_{i-1}+\xi_{1}+\eta)\nonumber\\
&\times Z_{(N-1),2(N-1)}(\{\lambda\}_{1\dots N-1},\{\xi\}_{2\dots N},\{\theta\})
\end{align}
\end{enumerate}
\begin{lemma}
The set of conditions i)-v) uniquely determines the partition function $Z_{N.2N}(\{\lambda\},\{\xi\},\{\theta\})$.
\end{lemma}
Indeed, it is sufficient to observe that the normalized partition function (\ref{normPartFun})
is a theta function of order $2N-2$ and norm $(N-1)\eta$ in each parameter $\lambda_{i}$. So we need $2N-1$ independent conditions to uniquely determine it. Using the symmetry ii) the recursion relations v) can be established for any point $\la_i=\xi_j$ ( or $\la_i=-\xi_j$ ). Hence we can prove by induction starting from the case $N=1$ that the partition function is uniquely determined as we need.
\begin{thm}
The partition function of the elliptic SOS model with reflecting end can be represented in the following form
\begin{align}\label{formula}
Z_{N,2N}&(\{\lambda\},\{\xi\},\{\theta\})=\gamma(\lambda)\det M_{i j}\prod_{i=1}^{N} \left(\frac{h(\theta+\eta(N-2i))}{h(\theta+\eta(N-2i+1))}\right)\nonumber\\
&\times\frac{\pl_{i,j=1}^{N}h(\lambda_{i}+\xi_{j})h(\lambda_{i}-\xi_{j})h(\lambda_{i}+\xi_{j}+\eta)h(\lambda_{i}-\xi_{j}+\eta)}{\pl_{1\leq i<j\leq N}h(\xi_{j}+\xi_{i})h(\xi_{j}-\xi_{i})h(\lambda_{j}-\lambda_{i})h(\lambda_{j}+\lambda_{i}+\eta)}
\end{align}
where the $N\times N $ matrix $M_{i j}$ can be expressed as
\begin{align}
M_{i,j}= &\frac{h(\theta+\zeta+\xi_j)}{h(\theta+\zeta+\lambda_{i})}\cdot\frac{h(\zeta-\xi_j)}
{h(\zeta+\lambda_{i})}\,\nonumber \\
\times &\frac{h(2\lambda_i)h(\eta)}{h(\lambda_{i}- \xi_{j}+\eta)h(\lambda_{i}+ \xi_{j}+\eta)h(\lambda_{i}-\xi_{j})h(\lambda_{i}+\xi_{j})}
\label{PFresult2}
\end{align}
\end{thm}
To prove the theorem, it is sufficient to check the property i) to v).

\begin{rem}
The partition function reduces to that of the trigonometric SOS model with reflecting end in the limit $p\rightarrow0$
\end{rem}
Taking the limit of  \eqref{formula} we obtain up to irrelevant numerical factor
\begin{align}
Z_{N,2N}&(\{\lambda\},\{\xi\},\{\theta\})=\gamma(\lambda)\det M_{i j}\prod_{i=1}^{N} \left(\frac{\sinh(\theta+\eta(N-2i)}{\sinh(\theta+\eta(N-i))}\right)\nonumber\\
&\times\frac{\pl_{i,j=1}^{N}\sinh(\lambda_{i}+\xi_{j})\sinh(\lambda_{i}-\xi_{j})\sinh(\lambda_{i}+\xi_{j}+\eta)\sinh(\lambda_{i}-\xi_{j}+\eta)}{\pl_{1\leq i<j\leq N}\sinh(\xi_{j}+\xi_{i})\sinh(\xi_{j}-\xi_{i})\sinh(\lambda_{j}-\lambda_{i})\sinh(\lambda_{j}+\lambda_{i}+\eta)}.
\end{align}
where the $N\times N $ matrix $M_{i j}$ is
\begin{align}
M_{i,j}= &\frac{\sinh(\theta+\zeta+\xi_j)}{\sinh(\theta+\zeta+\lambda_{i})}\cdot\frac{\sinh(\zeta-\xi_j)}
{\sinh(\zeta+\lambda_{i})}\,\nonumber \\
\times &\frac{\sinh(2\lambda_i)\sinh \eta }{\sinh(\lambda_{i}- \xi_{j}+\eta)\sinh(\lambda_{i}+ \xi_{j}+\eta)\sinh(\lambda_{i}-\xi_{j})\sinh(\lambda_{i}+\xi_{j})}.
\label{PFresult2}
\end{align}
We recover in this way the previous result of \cite{FilK10}.
\section*{Conclusions}
In this paper, we have computed the partition function of the elliptic SOS model with domain wall boundary conditions and one reflecting end. The main result is that this partition function is expressed as a single determinant, generalizing the recent observation in the trigonometric case that SOS type models with such boundaries give rise to a simpler representation of their partition functions. Recently, the trigonometric version of this SOS model was associated to open XXZ spin chains with particular boundary conditions in a very natural way by means of gauge transformation (Vertex-IRF, \cite{FilK}). It seems to be natural to ask for such relation between the result of this paper and the open XYZ spin chains, leading to the possibility to study XXZ and XYZ spin chains with boundary in a unified framework in opposition to the periodic case.
\section*{Aknowledgment}
The author is grateful to the LPTHE in Paris and IMB laboratory in Dijon for hospitality. I would like to thank N. Kitanine, V. Terras, J. Avan and V. Roubstov for many interesting discussions.


\begin{thebibliography}{10}

\bibitem{Al00}
T.D. Albert, H. Boos, R. Flume, R.H. Poghossian and K. Ruhlig,
\newblock Lett.Math.Phys.53 (2000) 201.

\bibitem{Bax73}
R.J. Baxter,
\newblock Ann.Phys 76 (1973), 1, 25, 48

\bibitem{Drin90}
V.Drinfel'd,
\newblock Len.Math. J. 1 (1990), 1419


\bibitem{Enr98}
B. Enriquez, G.Felder,
\newblock Comm.Math.Phys 195 (1998), 651

\bibitem{FadST79}
L.D. Faddeev, E.K. Sklyanin, L.A. Takhtajan,
\newblock Theor. Math. Phys. 40 (1979) 688.


\bibitem{Ize87}
A.G. Izergin,
\newblock Sov. Phys. Dokl. 32 (1987) 878.

\bibitem{IzeK84}
A.G. Izergin, V.E. Korepin,
\newblock Comm. Math. Phys. 94 (1984) 67.

\bibitem{KitMT99}
N. Kitanine, J.M. Maillet, V. Terras,
\newblock Nucl. Phys. B 554 [FS] (1999) 647, math-ph/9807020.

\bibitem{KitMT00}
N. Kitanine, J.M. Maillet, V. Terras,
\newblock Nucl. Phys. B 567 [FS] (2000) 554, math-ph/9907019.

\bibitem{KitKMNST07}
N. Kitanine et~al.,
\newblock J. Stat. Mech. Theory Exp.  (2007) P10009, 37 pp. (electronic).





\bibitem{KitKMNST08}
N. Kitanine et~al.,
\newblock J. Stat. Mech.: Theory Exp.  (2007) P07010,
\newblock arXiv:0803.3305.



\bibitem{Kor82}
V.E. Korepin,
\newblock Comm. Math. Phys. 86 (1982) 391.



\bibitem{Mai96}
J.M. Maillet, J. Sanchez de Santos,
\newblock in  L. D. Faddeev's Seminar on Mathematical Physics, 137178, Amer. Math. Soc. Transl. Ser. 2, 201, Amer. Math. Soc., Providence, RI, 2000.


\bibitem{Fel95}
G. Felder,
\newblock Proceedings of the {I}nternational {C}ongress of {M}athematicians,
  {V}ol.\ 1, 2 ({Z}\"urich, 1994), pp. 1247--1255, Basel, 1995, Birkh\"auser.

\bibitem{Fel06a}
G. Felder, A. Varchenko,
\newblock Comm. Math. Phys. 181 (1996) 741.

\bibitem{Fel06b}
G. Felder, A. Varchenko,
\newblock Nuclear Phys. B 480 (1996) 485.


\bibitem{FilK10}
G. Filali, N. Kitanine,
\newblock J.  Stat. Mech.: Theory  Exp. 2010 (2010)
  L06001, erratum:
  \newblock J. Stat. Mech.: Theory and Exp. 2010 (2010)
  E07002


\bibitem{FilK}
G. Filali, N. Kitanine,
\newblock arXiv:1011.0660v1 [math-ph]

\bibitem{JKMO88}
M. Jimbo, A. Kuniba, T. Miwa, M. Okado,
\newblock Commun. Math. Phys. 119 (1988), 543.


\bibitem{JKMO99}
M. Jimbo, H. Konno, S. Odake, J. Shiraishi,
\newblock Commun. Math. Phys. 199 (1999), 605.

\bibitem{Ros09}
H. Rosengren,
\newblock Adv. in Appl. Math. 43 (2009) 137.

\bibitem{PakRS08}
S. Pakuliak, V. Rubtsov, A. Silantyev,
\newblock J. Phys. A 41 (2008) 295204, 20.


\bibitem{Skl88}
E. Sklyanin,
\newblock J. Phys. A : Math. Gen. 21 (1988) 2375.

\bibitem{Weber}
H. Weber,
Elliptische Functionen und algebraische Zahlen, Friedrich Vieweg und Sohn, Braunschweig, 1891.

\bibitem{Whitaker}
E.T. Whittaker, G.N. Watson,
A Course of Modern Analysis, fourth ed., Cambridge University Press, Cambridge, 1927.


\end{thebibliography}
\end{document}